\documentclass[twocolumn, aps, amsmath,amssymb,showkeys,a4paper]{revtex4-1}%{revtex4-1}

\usepackage{amssymb}
\usepackage{amsbsy}
\usepackage{amsmath}
\usepackage{bbm}
\usepackage{xcolor}
\usepackage{graphicx}% Include figure files
\usepackage{dcolumn}% Align table columns on decimal point
\usepackage{bm}% bold math
\usepackage{subfigmat}

\begin{document}

%\preprint{APS/123-QED}

\title{Diffusiophoresis and medium structure control macroscopic particle transport in porous media}

\author{Mamta Jotkar}
\email{mamta.jotkar@upm.es, mamta.jotkar@idaea.csic.es }
\affiliation{Universidad Polit\'ecnica de Madrid, Spain, \\ Institute
  of Environmental Assessment and Water Research, Spanish National
  Research Council, Barcelona, Spain} 
\author{Pietro de Anna}
\email{pietro.deana@unil.ch}
\affiliation{Institute of Earth Sciences, University of Lausanne, Switzerland} 
\author{Marco Dentz}
\email{marco.dentz@idaea.csic.es }
\affiliation{Institute of Environmental Assessment and Water Research, Spanish National Research Council, Barcelona, Spain} 
\author{Luis Cueto-Felgueroso}
\email{luis.cueto@upm.es}
\affiliation{Universidad Polit\'ecnica de Madrid, Spain}%, ETSI Caminos, Canales y Puertos, Spain}
\date{\today}% It is always \today, today,

\begin{abstract}

In this letter, we show that pore-scale diffusiophoresis of colloidal particles along local
  salt gradients manifests in the macroscopic dispersion of particles
  in a porous medium. Despite is transient character, this microscopic phenomenon
  controls large-scale particle transport by altering their
  partitioning between transmitting and dead-end pores. It determines
  the distribution of residence and arrival times in the
  medium. Depending on the diffusiophoretic mobility, particles
  can be mobilized from or trapped in dead-end pores, which provides a
  means for the controlled manipulation of particles in porous media. 

\end{abstract}
\maketitle

Diffusiophoresis (DP) \cite{Derjaguin_1947} is the motion of
microscopic particles driven by local gradients of solute
concentration that has been demonstrated both
theoretically~\cite{PrieveAnderson_1984,Anderson_1989_ARFM,VelegolEtAl_2016,AultEtAl_2018}
and experimentally~\cite{AbecassisEtAl_2008,PalacciEtAl_2010_PRL,KarEtAl_2015,ShinEtAl_2016_PNAS,BattatEtAl_2019,SinghEtAl_2020_PRL}
to be a powerful particle manipulation tool. The physical mechanisms
that drive this complex physicochemical phenomenon can be broken down
to two components: chemiphoresis that occurs due the osmotic pressure
gradient along the surface of a charged particle (at the scale of the
particle) and electrophoresis arising due to the difference in the
diffusivities between the cation and the anion in the electrolyte
solution~\cite{VelegolEtAl_2016}. The physics behind this phenomenon
is well-established~\cite{PrieveAnderson_1984,Anderson_1989_ARFM}. Despite
the studies in relatively simpler microfluidic setups that demonstrate
particle focusing \cite{ShiEtAl_2016_PRL}, particle separation
\cite{ShinEtAl_2017_NatCommun,ShinEtAl_2018,RasmussenEtAl_2020,JotkarCueto-Felgueroso_2021},
particle banding \cite{StaffeldQuinn_1989}, particle trapping
\cite{SinghEtAl_2020_PRL}, etc., with the help of DP,
its effect on macroscopic transport within intricate and spatially
variable porous structures remains unexplored.

Flow and transport of dissolved solutes and suspended particles through
 porous and confined media are ubiquitous in natural and
engineered systems \cite{Bear}. Most geological and biological
porous media share the common feature of being spatially variable or
heterogeneous. The broad range of variability in pore size is known to
induce anomalous transport. {\color{black}Moreover,} the diversity in
shape of the constituent grains induces a rich flow organization that plays an
important role in groundwater contamination and
remediation~\cite{KahlerEtAl_2019_gw}, enhanced hydrocarbon
recovery~\cite{EOR}, transport through river sediments~\cite{Lei2022}
and water filtration systems~\cite{ShinEtAl_2017_NatCommun,miele2019}. {\color{black}The
  morphology of a porous medium is often characterized
  by the presence of cavities or dead-end pores (DEP), which represent
  the part of the system that cannot host net fluid transfer,
  resulting in stagnant flow~\cite{Lever1985, Nishiyama2017,
    Erktan2020}. These DEPs are connected via a network of
  percolating channels or transmitting pores (TP). Such DEP-TP}
structures characterize biological tissues~\cite{Hapfelmeier2010},
soil~\cite{Erktan2020} and filters~\cite{Phillip2011} and lead to
anomalous transport of passive tracers~\cite{BordoloiEtAl_2022}. {\color{black}To
date, the combined role of DP and the
complexity of the porous medium on particle transport remains elusive.} Here, we
use detailed numerical pore-scale simulations and analytical modeling to elucidate how
DP couples with the medium structure to alter macroscopic particles transport.

We study a {\color{black}fluid-}saturated porous system where a
particle suspension gets displaced by a continuously injected salt
solution. The velocity experienced by each transported 
particle results from advection in the flow field $\mathbf u$, which is
controlled by the medium structure and {\color{black} imposed flow
  rate}~\cite{Dentz2011,deAnnaPRF2017,Dentz2018}, and due to the
diffusiophoretic drift $\textbf{u}_{dp}$. In the thin Debye
layer limit, for dilute solutions with valence symmetric solutes
(e.g., LiCl, NaCl), the diffusiophoretic drift is proportional to the
{\color{black}gradient of the logarithm} of the solute concentration
\begin{equation}
	\label{udp}
	\textbf{u}_{dp} = \Gamma_p \nabla \ln s
\end{equation}
and the diffusiophoretic mobility $\Gamma_p$ is approximately a
constant. $\Gamma_p$ is determined by the size and surface charge of
the particle \cite{PrieveAnderson_1984,Anderson_1989_ARFM} 
\begin{align}
\nonumber
	\Gamma_p = \frac{\varepsilon k_B T}{\nu Z e } \left\{
          D\zeta - \frac{2k_B T}{Z e} \ln \left[1 - \tanh^2
            \left(\frac{Z e \zeta}{4k_B T}\right) \right]\right\},
\end{align}
where $\varepsilon$ is the dielectric permittivity of the medium,
$\zeta$ is the particle zeta potential, $\nu$ is
the kinematic viscosity of the medium, $k_B$ is the Boltzmann
constant, $T$ is the absolute temperature, $Z$ is the valence of the
constituent ions of the solute, $e$ is the proton charge and
$D=(D_+ - D_-)/(D_+ + D_-)$ measures the difference in
diffusivity $D_+$ of the cation and $D_-$ of the anion. The
logarithmic dependence of the particle velocity $\textbf{u}_{dp}$ on
the solute concentration gradients $\nabla s$ allows for rapid and
efficient particle motion, even in low concentration areas.

\begin{figure}[t]
 \begin{center}
 \includegraphics[width=0.45\textwidth]{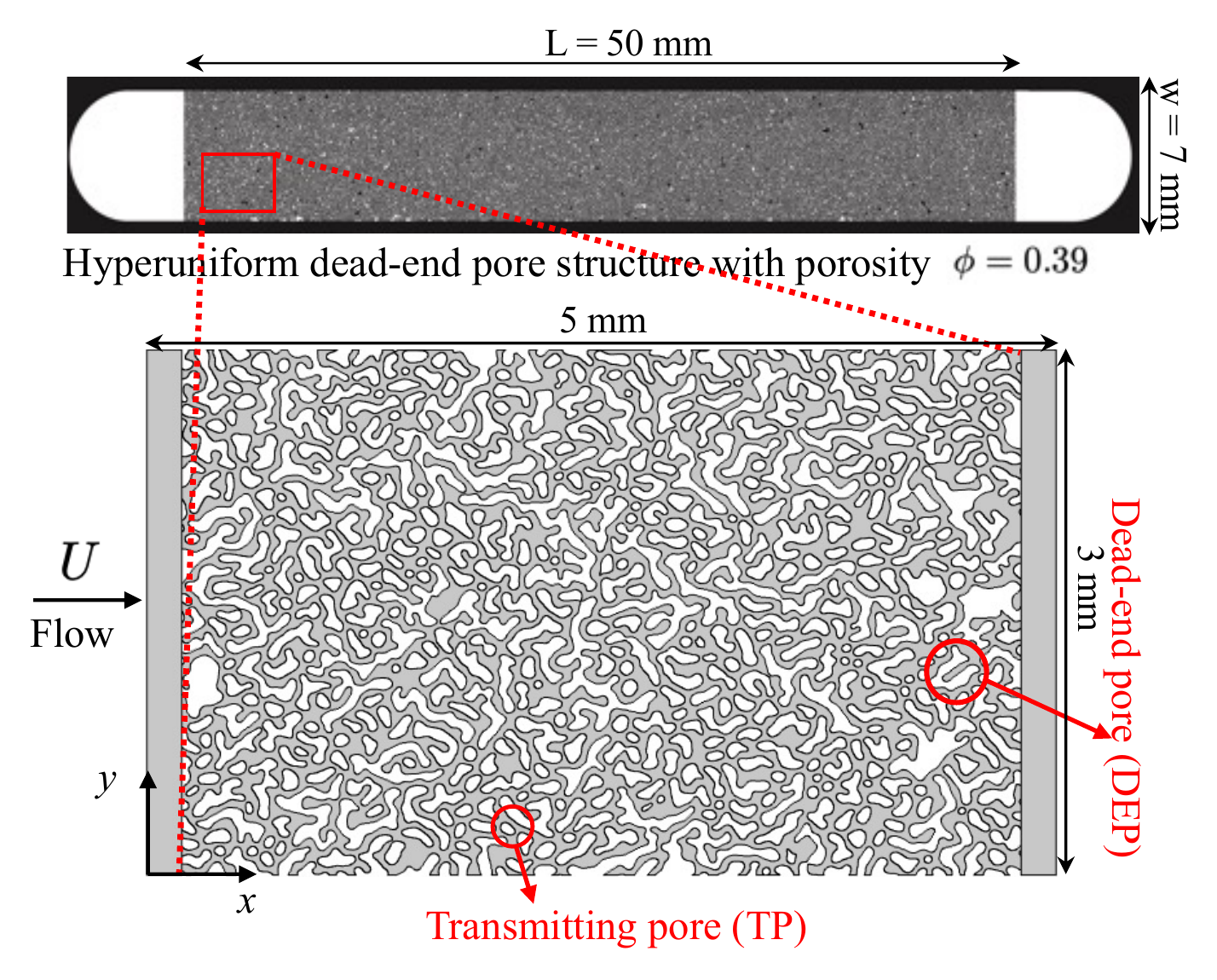}
 \vspace{-0.4cm}
 \caption{Hyper-uniform porous structure characterized by DEPs and TPs. Computational domain with white spaces indicating solid grains (bottom). \label{fig1}}
 \end{center}
 \end{figure}

We consider a porous system {\color{black}characterized by DEPs of
  different size connected to a network of TPs,}
similar to the one used in ref.~\cite{BordoloiEtAl_2022}. The computational
domain is shown in figure \ref{fig1}. {\color{black} The mean flow is
  driven from left to right with the flow rate $U$.} The medium is statistically homogeneous with a mean
pore-size $\lambda=30 \mu \text{m}$ and porosity $\phi= 0.39$. The dual
feature of this medium is characterized by TPs, and DEPs leading to stagnant flow. While solutes typically diffuse and dissipate
gradients within pores over time scales shorter than the time $\tau_L = L/U$
needed to elute a pore-volume, the junctures of TPs and DEPs serve as excellent
candidates to retain gradients of solute concentration, which in turn trigger
DP.

In the low Reynolds number limit, the
{\color{black}fluid-solute-particle dynamics} are governed by the Stokes
equation for fluid flow, and the advection-diffusion equations for the
solute and particle transport \cite{AultEtAl_2018}
\begin{subequations}
	\label{govern_eq}
	\vspace{-0.cm}
	\begin{equation}
		\label{nse}
		0 = \frac{1}{\rho} \left( - \nabla p  +  \mu{\nabla}^2 \textbf{u} \right), 
	\end{equation} \vspace{-0.8cm}
	\begin{equation}\label{cont}
		\nabla  \cdot \textbf{u} =0, 
	\end{equation}  \vspace{-0.9cm}
	\begin{equation}\label{scon}
		 \frac{\partial s}{\partial t} +  \nabla  \cdot (\textbf{u} s )=  D_s{\nabla}^2 s , 
	\end{equation} \vspace{-0.7cm}
	\begin{equation}\label{ccon}
		 \frac{\partial c}{\partial t} +  \nabla  \cdot (\textbf{u} +\textbf{u}_{dp})c =  D_p{\nabla}^2 c ,
	\end{equation} 
\end{subequations}
where $\textbf{u}$ is the two-dimensional velocity field, $p$ is the
pressure, $s$ is the solute concentration and $c$ is the 
particle concentration, $\mu$ is the dynamic viscosity of the fluid
and $\rho$ its density. $D_s$ and $D_p$ are the diffusion coefficient
of solute and particles, respectively. Typically, solutes diffuse much
faster than particles such that $D_s \gg D_p$. Solute and particle
transport can be characterized by the respective P\'eclet numbers,
$Pe_s = U \lambda / D_s $ and $Pe_p = U \lambda / D_p$, which compare
the characteristic diffusion time scales $\tau_{D_s} = \lambda^2 /
D_s$ and $\tau_{D_p} = \lambda^2 /D_p$ to the advection time $\tau_v =
\lambda / U$ over the {\color{black} mean pore length}. The
diffusiophoretic particle velocity $\textbf{u}_{dp}$ is given by
Eq.\eqref{udp}. The medium is initially saturated with particles and
solute with initial concentrations $c_i$ and $s_i$, respectively. At time $t> 0$, 
 a sharp front of solute at concentration
 $s_H \gg s_i$ is injected such that the ratio $\chi = s_i/s_H \ll
 1$. This solute front induces a dynamic and heterogeneous solute concentration gradient
that drives DP. The governing
equations~\eqref{govern_eq} are solved numerically for the pore
geometry detailed above. The numerical setup is described
in~\cite{SM}. 

%%%%%%%%%%%%%%%%%%%%%%%%%%%%%%%%%%%%%%%%%%%%%%%%%%%%%%
 \begin{figure*}[t]
 \begin{center}
 $t\sim35\tau_v$ \hspace{2.6cm} $t\sim70\tau_v$ \hspace{2.6cm} $t\sim210\tau_v$ \hspace{2.6cm} $t\sim665\tau_v$
 \subfigure[Particle trapping $\Gamma_p<0$]{\label{fig2a}\includegraphics[trim=0.9cm 1.9cm 3.5cm 2.9cm, clip=true, width=1.7in]{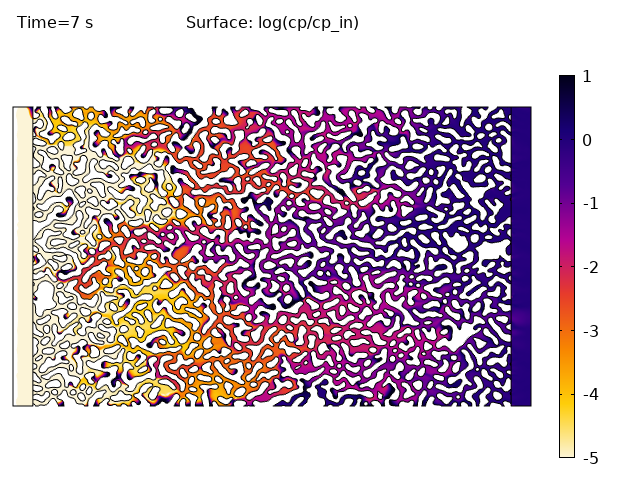}
 \includegraphics[trim=0.9cm 1.9cm 3.5cm 2.9cm, clip=true, width=1.7in]{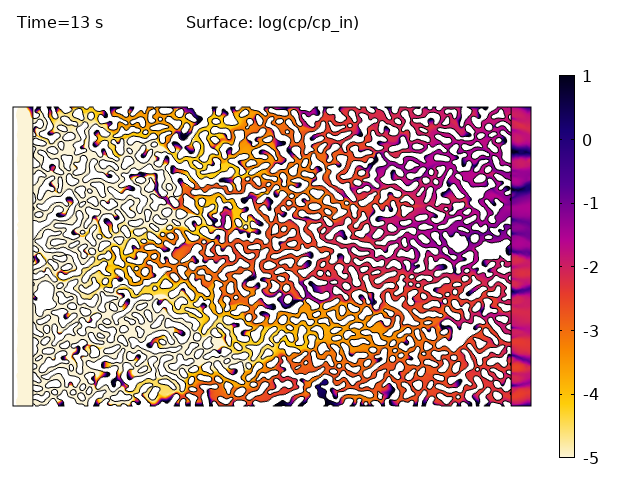}
 \includegraphics[trim=0.9cm 1.9cm 3.5cm 2.9cm, clip=true, width=1.7in]{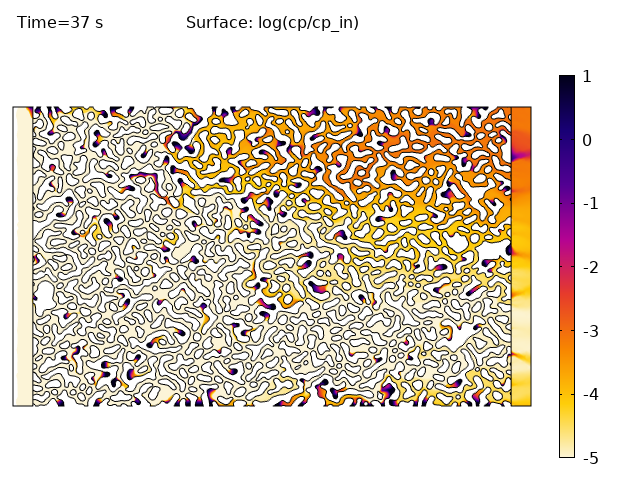}
 \includegraphics[trim=0.9cm 1.9cm 3.5cm 2.9cm, clip=true, width=1.7in]{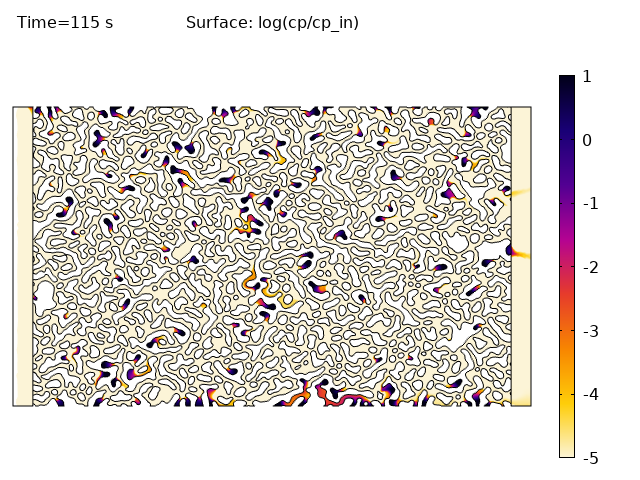}
 } \\ 
 \subfigure[No DP $\Gamma_p=0$]{\label{fig2b}
 \includegraphics[trim=0.9cm 1.9cm 3.5cm 2.9cm, clip=true, width=1.7in]{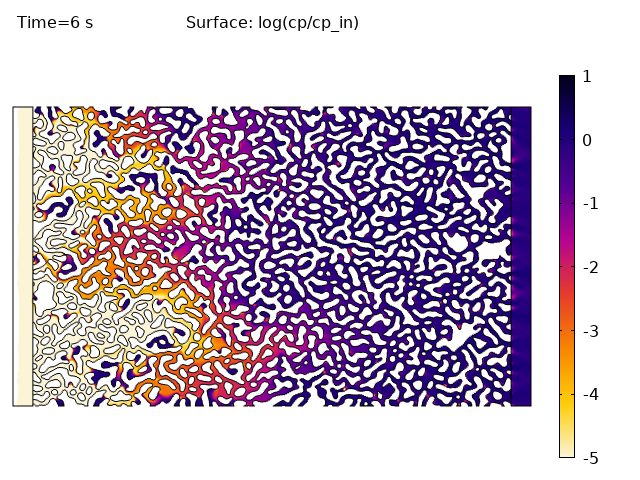}
 \includegraphics[trim=0.9cm 1.9cm 3.5cm 2.9cm, clip=true, width=1.7in]{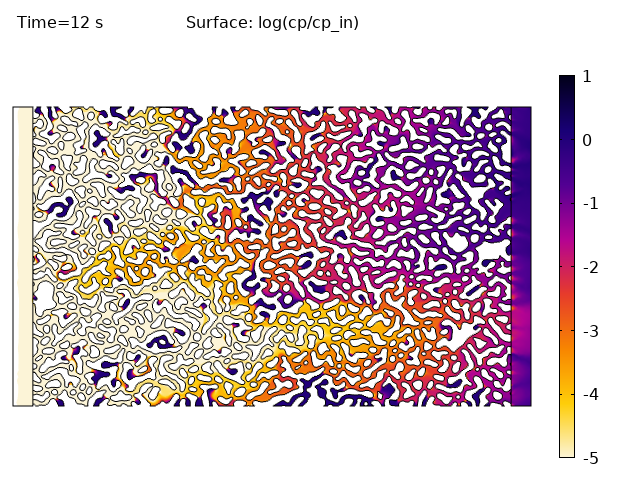}
 \includegraphics[trim=0.9cm 1.9cm 3.5cm 2.9cm, clip=true, width=1.7in]{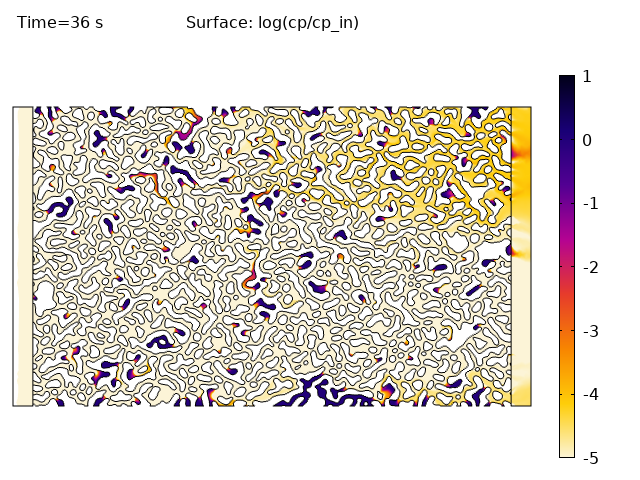}
 \includegraphics[trim=0.9cm 1.9cm 3.5cm 2.9cm, clip=true, width=1.7in]{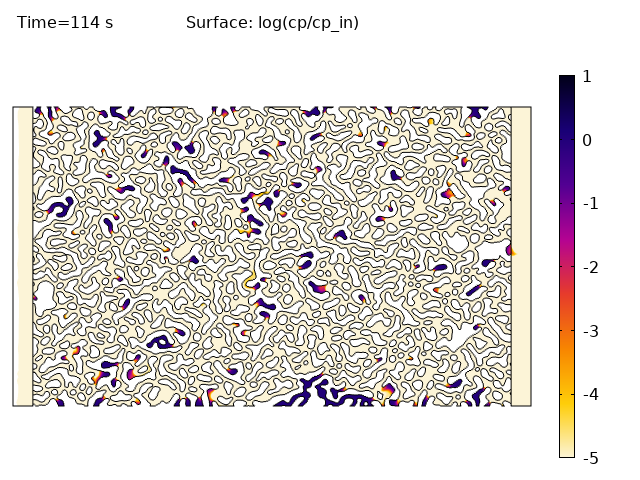}
 }\\
 \subfigure[Particle extraction $\Gamma_p>0$]{\label{fig2c}
 \includegraphics[trim=0.9cm 1.9cm 3.5cm 2.9cm, clip=true, width=1.7in]{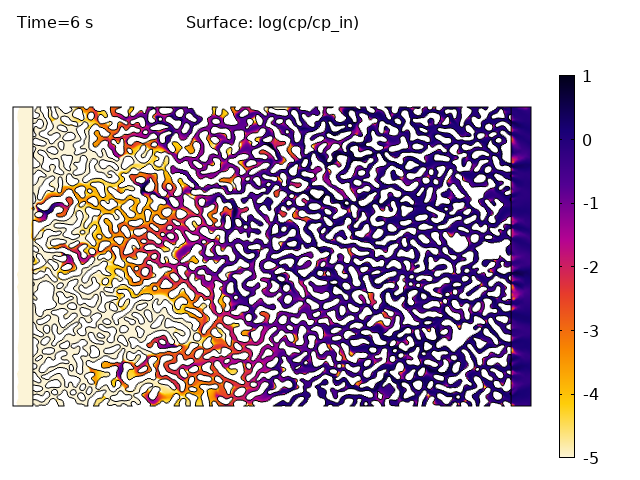}
 \includegraphics[trim=0.9cm 1.9cm 3.5cm 2.9cm, clip=true, width=1.7in]{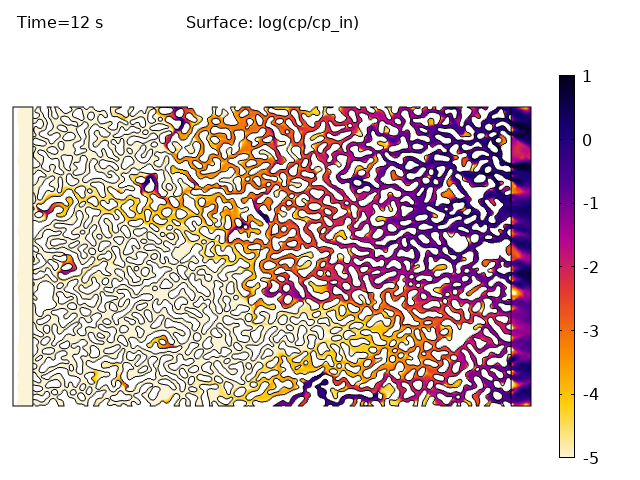}
 \includegraphics[trim=0.9cm 1.9cm 3.5cm 2.9cm, clip=true, width=1.7in]{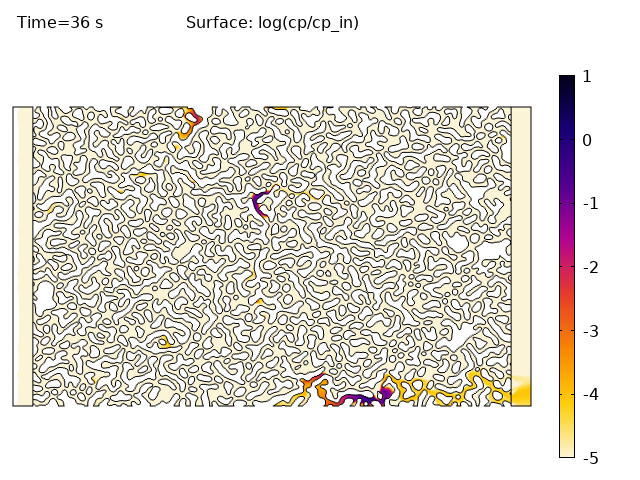}
 \includegraphics[trim=0.9cm 1.9cm 3.5cm 2.9cm, clip=true, width=1.7in]{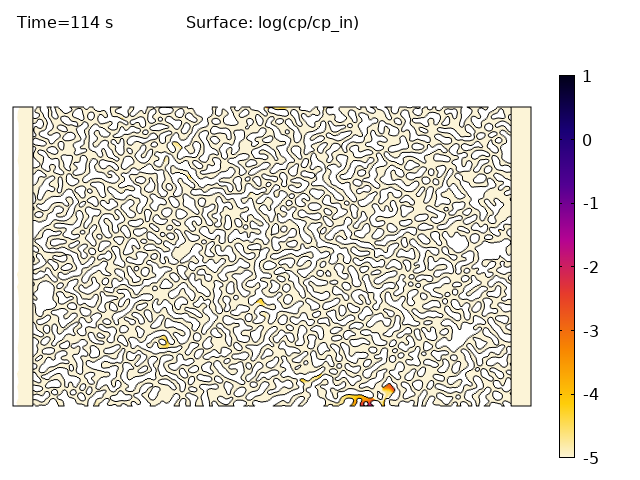}
 }
 \vspace{-0.3cm}
 \caption{Temporal evolution of the dimensionless particle
   distributions $c/c_i$ for (a) trapping, (b) no-DP, and (c)
   extraction cases. The concentration scale is logarithmic between $10^{-5}$
   (light) and $10$ (dark). White spaces indicate solid
   grains. \label{fig2}}
 \end{center}
 \end{figure*}
%%%%%%%%%%%%%%%%%%%%%%%%%%%%%%%%%%%%%%%%%%%%%%%%%%%%%%

Figure \ref{fig2} shows the temporal evolution of the particle
concentration field without DP ($\Gamma_p=0$), and for {\color{black}diffusiophoretic trapping ($\Gamma_p<0$), and extraction ($\Gamma_p>0$)}. In the absence of DP, majority of the particles get dispersed through the
TPs leaving behind a small fraction of particles that accumulate
within the DEPs, where the flow is stagnant, 
organized in convection rolls~\cite[][]{BordoloiEtAl_2022}. The only mechanism
through which these localized particles can escape into the TPs
is diffusion, the times scale of which is typically orders of
magnitude {\color{black} larger} than $\tau_L$. 

The injection of a sharp front of solute at a higher concentration
results in local gradients of solute concentration that 
drive DP within the DEPs. For $\Gamma_p < 0$, that is, when particles
migrate from high to low solute concentrations, DP leads
to the trapping of particles inside the DEPs, as shown in figure \ref{fig2a}.
% {\color{black} Note that this scenario is analogous to reversing the
% direction of the solute gradient when the displacing front carries
% solute at lower concentration ($\chi=s_i/s_H\gg 1$). (MAMTA:
% Something like this sentence is needed to give context to the 1st
% line in the very last paragraph about reversibility)}

\begin{figure}[t]
 \begin{center}
 \includegraphics[width=0.45\textwidth]{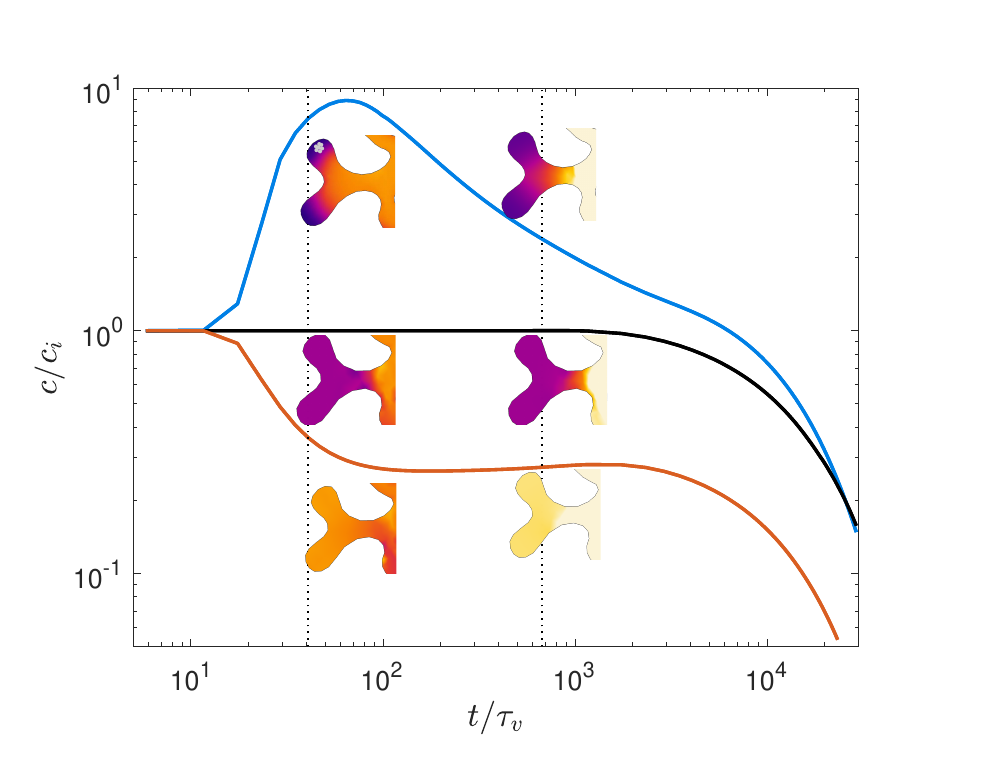}
 \caption{Temporal evolution of dimensionless particle concentration
   ($c/c_i$) at an arbitrary point (indicated by grey) within a DEP
   shown in the inset for {\color{black}$\Gamma_p^*=-1$ (blue, trapping), $\Gamma_p^*=0$
   (black, no DP), and $\Gamma_p^*=0.4$} (red, extraction). The particle
   concentrations shown in the inset are displayed in a logarithmic scale and range 
   from $10^{-6}$ (light) to $10$ (dark).
   \label{fig3}}
 \end{center}
 \end{figure}

When $\Gamma_p>0$, DP leads to 
particle mobilization out of the DEPs because the particles
move towards the higher solute concentration zones in the TPs. This is
seen in the particle distributions shown in Figure \ref{fig2c}, where the
particles within the DEPs rapidly escape the DEPs and leave
the domain from the right outlet. Figure~\ref{fig3} shows the time evolution of
the solute concentration at a point within a DEP {\color{black}close to the bottom of the DEP}. The increase or
decrease of concentration due to DP occurs on the time scale $\tau_{D_s}${\color{black}, which here is smaller than $\tau_{D_p}$ by a factor of $10^3$.} Thus, the action of DP on mass
transfer between TPs and DEPs is limited to
relatively short initial times.

To understand how DP controls the macroscopic fate of the suspended particles, 
we first estimate the fraction $\alpha$ of particles that are trapped in or
mobilized from the DEPs in the initial phase. 
To this end, we consider a single DEP connected to a TP~\cite{SM}. For this geometry, we can derive the solute concentration profile in the DEP, and
thus obtain explicit expressions for the diffusiophoretic drift $u_{dp}$. Combining the latter with conservation of particle flux at the interface between TP and DEP, we obtain the following expression for $\alpha$ as
a function of $\Gamma_p^\ast = \Gamma_p / U \lambda$ 
\begin{widetext}
\begin{align}
\alpha = \alpha _0 \left[1 - \Gamma_p^\ast (1-\chi)Pe_s\right] +
  H(-\Gamma_p^\ast) \left\{ \frac{{2 \Gamma_p^\ast}^2
  (1 - \chi)^2 Pe_s Pe_{p} \ell_0^\ast}{\pi} 
\ln\left[\frac{2\Gamma_p^\ast (1 - \chi) Pe_{p} \ell_0^\ast}{2\Gamma_p^\ast (1 - \chi) Pe_{p} \ell_0^\ast-  \pi
 } \right] \right\},
	\label{alpha_mp}
\end{align}
\end{widetext}
where $H(\cdot)$ is the Heaviside step function, and $\alpha_0$ the initial
fraction of particles in the DEPs {\color{black}in the absence of DP
  ($\Gamma_p=0$)}. The dimensionless length $\ell_0^\ast$ denotes 
a characteristic diffusion scale at the interface between TP and DEP.
It is inversely proportional to the particle P\'eclet
number, $\ell_0^\ast \sim 1/Pe_p$. We set in the following $\ell_0^\ast =
\beta/Pe_p$ with $\beta$ a number of the order of one. Expression~\eqref{alpha_mp} predicts that the
particles are depleted from the DEPs for $\Gamma_p^\ast \geq 1/(1 - \chi)
Pe_s$. Furthermore, for strongly negative diffusiophoretic mobility, the rate at which particles are transferred to
the interface is eventually smaller than the diffusiophoretic
drift. Thus, {\color{black} by taking the limit of $\Gamma_p^\ast \to -
  \infty$ in Eq. (\ref{alpha_mp}),} we find that $\alpha$ asymptotes toward
\begin{align}
\alpha_{\infty} = \alpha_0 \left(1 + \frac{\pi Pe_s}{\beta}\right).
\end{align}
The trapped particle fraction increases linearly with $Pe_s$ because
the solute gradients and thus the diffusiophoretic drift increase
with increasing $Pe_s$. 

%%%%%%%%%%%%%%%%%%%%%%%%%%%%%%%%%%%%%%%%%%%%%%%%%%%%%%
%%%%%%%%%%%%%%%%%%%%%%%%%%%%%%%%%%%%%%%%%%%%%%%%%%%%%%

%%%%%%%%%%%%%%%%%%%%%%%%%%%%%%%%%%%%%%%%%%%%%%%%%%%%%%
 \begin{figure}[t]
 \begin{center}
 \includegraphics[width=0.45\textwidth]{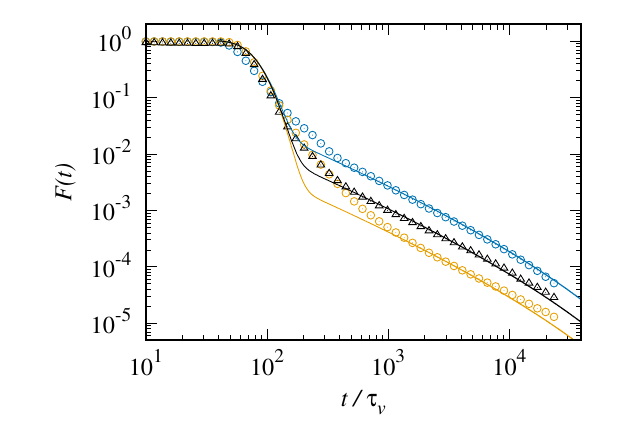} 
 \includegraphics[width=0.45\textwidth]{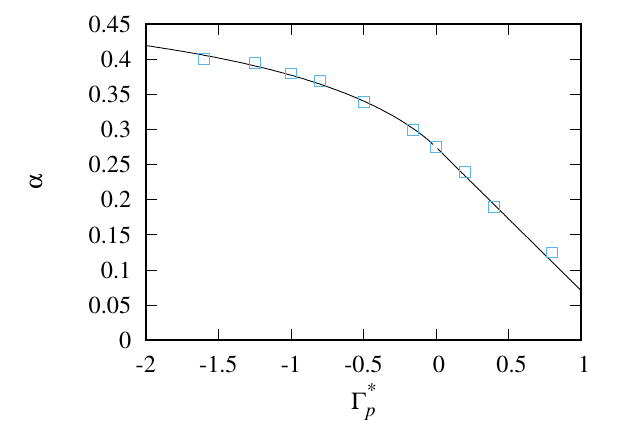} 
 \caption{(Top panel) Arrival time distribution $F(t)$ at the outlet
   for (blue circles)  {\color{black}$\Gamma_p^*= -1$, (black triangles) $0$ and
   (orange circles) $0.4$}. The solid lines denote the travel-time model in
   Eq.~\eqref{btc_ctrw}. (Bottom panel) Particle fraction $\alpha$ in
   the DEP from the (symbols) numerical data, and (solid line) the
   analytical expression~\eqref{alpha_mp} for $\ell_0^\ast = 0.65/Pe_p$. \label{fig4}}
 \end{center}
 \end{figure}
%%%%%%%%%%%%%%%%%%%%%%%%%%%%%%%%%%%%%%%%%%%%%%%%%%%%%%

Figure \ref{fig4} shows the distribution of arrival times of 
particles at the outlet. Similar to Ref.~\cite{BordoloiEtAl_2022}, we observe two transport
regimes. At times of the order of $\tau_L$, particles at the outlet
are produced by advection and dispersion from the TPs.
{\color{black}For times $t \gg \tau_L$, the arrival time distribution
  deviates from the exponential decay predicted under the classical
  dispersion framework and displays a power law tail. This is caused
  by the particles that are initially trapped within the DEPs.} For $\Gamma_p < 0$, {\color{black} an
  increasing fraction of} particles is trapped {\color{black}in the DEP}, which manifests in a stronger tail than without DP. 
 For $\Gamma_p > 0$, particles are  extracted from the DEP at initial
 times, and thus the tail is weaker than for $\Gamma_p \leq 0$.

The arrival time distribution can be modeled as
the superposition of the residence time distributions in the TPs and
DEPs~\cite{BordoloiEtAl_2022}
\begin{equation}
F(t) = (1-\alpha) % \frac{1}{L} \int_0^L \text{d}x f_0(t,x)
F_0(t)+ \alpha \int_0^\infty \text{d}\tau \frac{g(t/\tau)}{\tau} f_D(\tau),
\label{btc_ctrw}
\end{equation} 
respectively. Note that $\alpha$ is the fraction of trapped  particles
after the short initial phase, which can be approximated by
expression~\eqref{alpha_mp}.
We assume that transport in the TP can be
characterized by the mean flow velocity and a hydrodynamic dispersion
coefficient. Thus, $F_0(t)$ is given by the superposition of inverse
Gaussian distributions as, 
\begin{align}
F_0(t) = \frac{1}{L} \int\limits_0^L d x \frac{x \exp[-(x - \overline u t)^2/(4 D_h t)]}{\sqrt{4 \pi D_h t^3}}. 
\end{align}
%
%MARCO: What value was used for $D_h$? MAMTA: $D_h=0.1mm^2/s$
As we see in Figure~\ref{fig4}, this approximation is able to capture the early
arrival times, but underestimates the arrival time distribution at intermediate
times. This can be traced back to the velocity variability between
pores~\cite{Dentz2018}. The residence time distribution within the DEPs
is expressed in terms of {\color{black}a Gamma-}distribution $g(t')$ of dimensionless
residence times $\tau' = \tau/\tau_{D_p}$ in a single DEP and the
distribution $f_D$ of characteristic diffusion times $\tau_{D_p}$, {\color{black}given below}~\cite[][]{BordoloiEtAl_2022}
\begin{equation}\label{pdf_dep}
	g(t') = \frac{t'^{-2/3}\exp (-t')}{\Gamma (1/3)},
\end{equation}
while $f_D$ is obtained from the distribution $f_\Lambda$ of aspect ratios
$\Lambda$ through the map $\Lambda \to \tau_{D_p}  = (\lambda\Lambda)^2
/D_p$.

{\color{black}Figure \ref{fig4} shows the numerically estimated
  particle fractions versus the dimensionless diffusiophoretic
  mobility, and the analytical expression~\eqref{alpha_mp} for $\beta
  = 0.65$.}  
The analytical model describes the full dependence of $\alpha$ on $\Gamma_p$ for
both extraction from and trapping in DEPs, and thus seems to capture the controls of
DP on the macro-scale dispersion of  particles.

In conclusion, the microscopic interactions between DP and flow and transport
through porous media impact the macroscopic fate of 
particles. Depending on the diffusiophoretic mobility
$\Gamma_p$, DP may promote trapping {\color{black}within} or it may
lead to particle mobilization from the DEPs. This can be exploited
for preferential particle deposition or removal. DP is a short-term phenomenon
that persists as long as the solute gradients are not dissipated by diffusion,
which is characterized by the time scale $\tau_{D_s} = \lambda^2/D_s$ {\color{black} and is typically smaller than the characteristic
advection time across the medium i.e. $\tau_{D_s} \ll \tau_L$.} {\color{black} However, despite its short
  timespan, DP has a significant impact on macroscopic
  transport, by reorganizing the partitioning of particles between the
  DEPs and TPs, which is quantified by the fraction $\alpha$.}

Our results suggest that DP provides an {\color{black}efficient} way of
controlling particle transport through porous media in a reversible
manner by changing the direction of the solute gradient. Moreover, the existence of
DEPs is quite common in natural geological and biological porous
media, and serve as excellent candidates for retaining microscopic gradients of
solute concentration for relatively large times. This opens new
avenues for developing technological solutions to
various problems of socio-economic relevance such as groundwater
remediation, enhanced oil recovery, water-filtration systems, targeted drug
delivery and microfluidics for biomedical applications.

\begin{acknowledgments}
	This work has received funding from European Union's Horizon 2020
        research and innovation program under the Marie Sk\l{}odowska-Curie
        G.A. No. 895569. MD and LCF gratefully acknowledge funding from the
        Spanish Ministry of Science and Innovation through the project HydroPore
        (PID2019-106887GB-C31-C33). P.d.A. acknowledges the support of FET-Open
        project NARCISO (ID: 828890) and of Swiss National Science Foundation
        (grant ID 200021 172587).
       
\end{acknowledgments}

%\bibliographystyle{unsrt}
%\bibliography{ref}

\end{document}

%\preprint{APS/123-QED}

\title{Supplemental Material: Diffusiophoresis and medium structure control macroscopic particle transport in porous media}

\author{Mamta Jotkar}
\email{mamta.jotkar@upm.es, mamta.jotkar@idaea.csic.es }
\affiliation{Universidad Polit\'ecnica de Madrid, Spain, \\ Institute of Environmental Assessment and Water Research, Spanish National Research Council, Barcelona, Spain} 
\author{Pietro de Anna}
\email{pietro.deana@unil.ch}
\affiliation{Institute of Earth Sciences, University of Lausanne, Switzerland} 
\author{Marco Dentz}
\email{marco.dentz@idaea.csic.es }
\affiliation{Institute of Environmental Assessment and Water Research, Spanish National Research Council, Barcelona, Spain} 
\author{Luis Cueto-Felgueroso}
\email{luis.cueto@upm.es}
\affiliation{Universidad Polit\'ecnica de Madrid, Spain}%, ETSI Caminos, Canales y Puertos, Spain}
\date{\today}% It is always \today, today,

\begin{abstract}

This supplemental material gives details on the setup of the detailed
numerical simulations of flow and solute and particle transport, and
the derivation of the one-dimensional analytical model for the
quantification of the fraction of particles trapped in the DEPs. 

\end{abstract}

\date{\today}

\maketitle

% %%%%%%%%%%%%%%%%%%%%%%%%%%%%%%%%%%%%%%%%5
% \begin{figure}[h]
% \begin{center}
% \includegraphics[width=.6\textwidth]{./Figure/setup}
% \caption{Hyper-uniform porous structure characterised by dead-end
%   pores (DEPs) and transmitting pores (TPs). Computational domain with white spaces
%   indicating solid grains (bottom). \label{fig1}}
% \end{center}
% \end{figure}
% %%%%%%%%%%%%%%%%%%%%%%%%%%%%%%%%%%%%%%%%5

\section{Numerical simulations}
\begin{figure}[h]
\begin{center}
\includegraphics[width=0.7\textwidth]{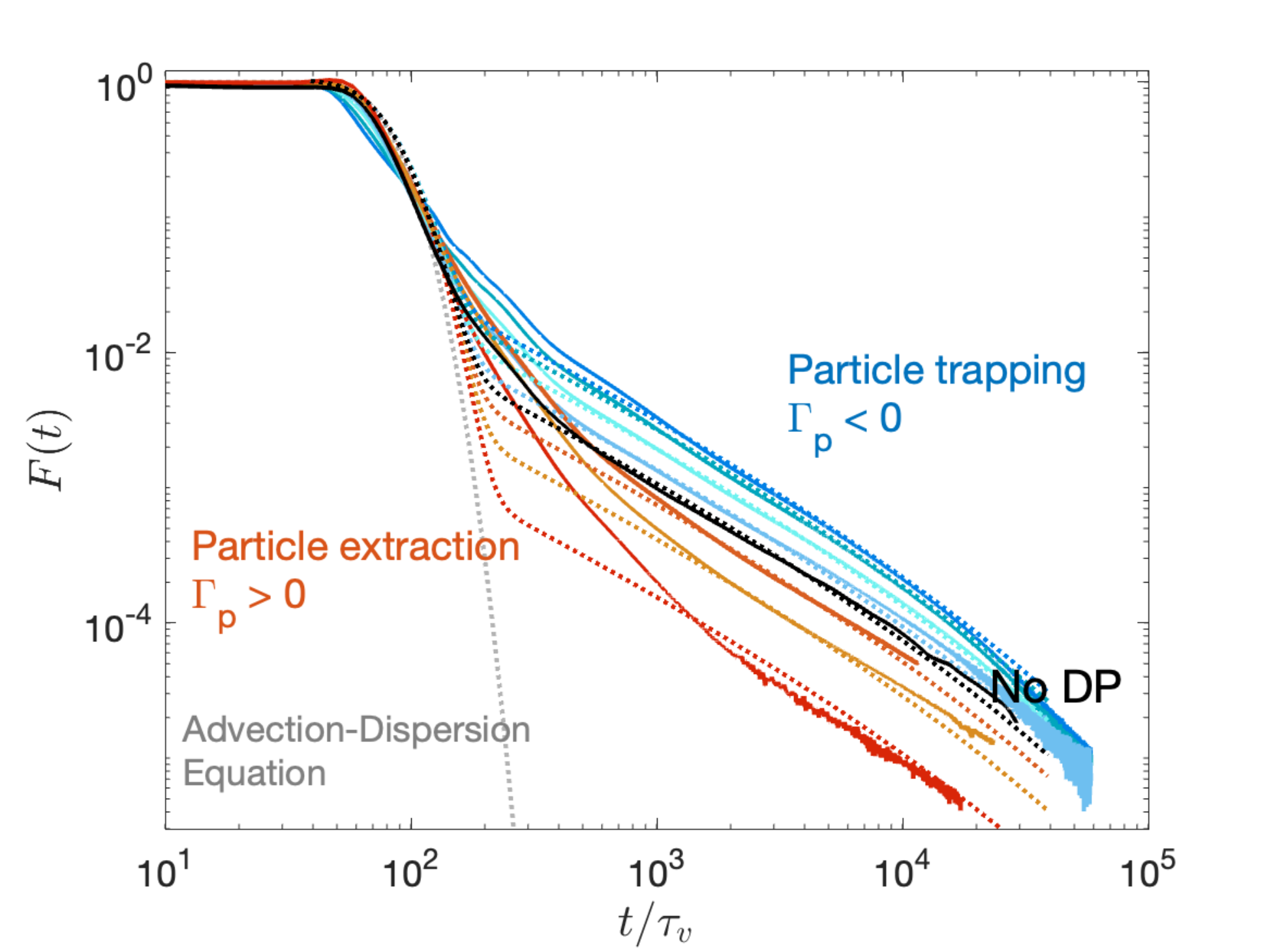}
\caption{Numerical simulations. Arrival time distribution at the outlet for different diffusiophoretic mobilities
  $\Gamma_p/\lambda U$ ranging from -1.6, -1, -0.5, -0.01 (shades of
  blue, trapping) to 0 (black, no DP) to 0.02, 0.4, 0.8 (shades of red, extraction). Solid curves correspond to
  numerical simulations while dotted curves correspond to the
  one-dimensional travel-time model obtained using
  different fraction $\alpha$ of particles initially available within
  the DEPs. \label{fig2}}
\end{center}
\end{figure}

The computational domain is shown in figure 1 in the main text. The mean flow is
from left to right. This medium is statistically homogeneous such that
the distribution of the pore-size is narrow with a strong peak close
to mean pore-size $\lambda=30 \mu \text{m}$. It exhibits complex pore
network interspersed among disordered solid grains with a porosity
$\phi= 0.39$. The domain is initially saturated with solute at a lower concentration ($s_i=$0.1mM) and particles. A sharp
front of solute at a higher concentration ($s_H=$10mM) is then injected such that the
ratio $\chi=s_i/s_H=0.01$. The governing
equations (Eq.(2) in the article) are subjected to no slip and no
penetration flow at the solid surfaces, uniform flow with an average
fluid speed $U$ at the left inlet and constant pressure at the right
outlet. For the solute concentration, we use constant flux at the
inlet and no flux boundary condition everywhere else whereas, for the
particle concentration, we impose a conservative form of no flux
boundary condition, where the sum of diffusive and advective fluxes is
zero. As an initial condition, we assume initially no flow $\textbf{u}
= 0, p = 0$ and impose solute concentration $s_i = 0.1$mM and particle
concentration $c_i=0.1$mM. The numerical model has been validated for
a simpler micro-channel geometry \cite{JotkarCueto-Felgueroso_2021} as
well as for the hyper-uniform porous medium considered here by
approximating the case without DP with
Ref. \cite{BordoloiEtAl_2022}. We use COMSOL
Multiphysics\textregistered $\;$based on finite element for performing
the pore-scale simulations \cite{comsol}. For the physical parameters
we use the following values (closest to realistic ones): $\mu=10^{-3}
Pa.s$, $\rho=10^3 kg/m^3$, $D_s = 7 \times 10^{3}\mu \text{m}^2/s$,
$D_p = 7  \mu \text{m}^2/s$ and $U=175 \mu \text{m}/s$. This yields
a characteristic advection time of $\tau_v = 0.17$ s, the salt
diffusion time $\tau_{D_s} = 0.12$ s, the particle diffusion time
$\tau_{D_p} = 128$ s, and the   P\'eclet numbers $Pe_s = U \lambda /
D_s = 0.75$ based on the solute and $Pe_p = U \lambda / D_p = 750$
based on the particles. Particle concentrations at the outlet for different mobilities
$\Gamma_p$ are shown in figure \ref{fig2}. 

%%%%%%%%%%%%%%%%%%%%%%%%%%%%%%%%%%%%%%%%5
%%%%%%%%%%%%%%%%%%%%%%%%%%%%%%%%%%%%%%%%5

%%%%%%%%%%%%%%%%%%%%%%%%%%%%%%%%%%%%%%%%5
\section{Analytical model }
%%%%%%%%%%%%%%%%%%%%%%%%%%%%%%%%%%%%%%%%5

%%%%%%%%%%%%%%%%%%%%%%%%%%%%%%%%%%%%%%%%5
\begin{figure}[h]
\begin{center}
\includegraphics[width=0.7\textwidth]{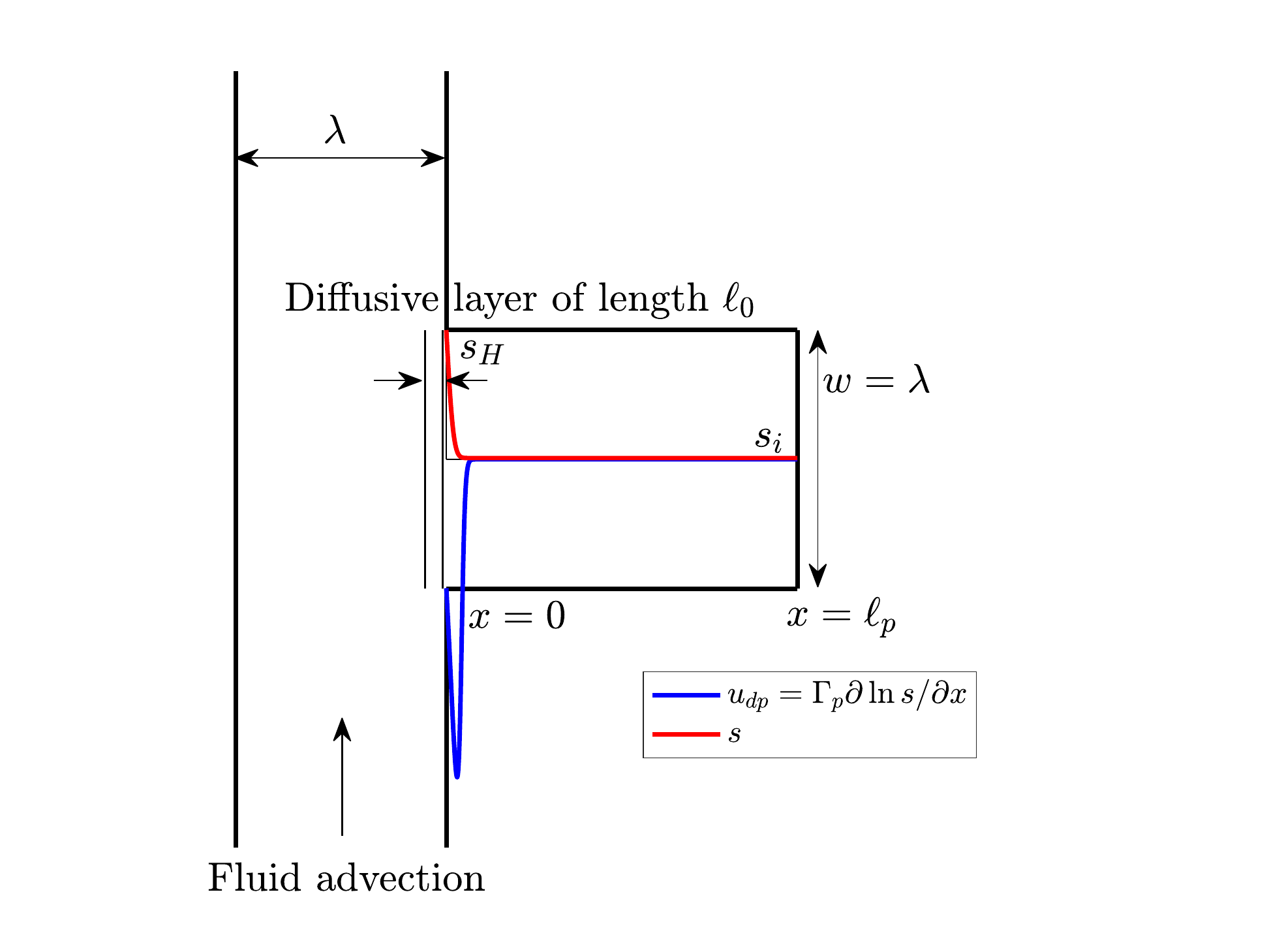}
\caption{Illustration of the one-dimensional analytical model. A single DEP of length $\ell_p$ is  connected to a (vertical) TP with a
  width of $\lambda$. The solute concentration within the DEP is
  illustrated by the red line, the blue line denotes the profile of $u_{dp}$. At short times, 
the diffusiophoretic drift is strongly localized at the interface between TP and DEP.   \label{1d_model}}
\end{center}
\end{figure}
%%%%%%%%%%%%%%%%%%%%%%%%%%%%%%%%%%%%%%%%5

We construct a one-dimensional model to quantify the dependence of the initial fraction of
particles $\alpha$ within the DEPs on the DP mobility $\Gamma_p$ (see
figure \ref{1d_model}). To this end, we assume that particle transport in
the DEP at short times is dominated by the diffusiophoretic drift such that
\begin{align}
\label{eq:adv}
\frac{\partial c}{\partial t} + \frac{\partial}{\partial x} u_{dp} c = 0. 
\end{align}
The total mass of particles in the DEP is given by
\begin{align}
\label{mint}
m_{dp} = w \int\limits_0^{\ell_p} dx c,
\end{align}
where $w$ is the pore-width. Since the only flux of particles
  toward or from the DEP is across  the DEP-TP junction, the temporal
  variability of the mass of particles in the DEP is equal to the mass flux at $x =
  0$ and controlled by DP. Spatial integration of
  Eq.~\eqref{eq:adv} according to Eq.~\eqref{mint} gives
\begin{equation}
\label{eq:mdp}
\frac{\partial m_{dp}}{\partial t} = w u_{dp}(x=0,t) c(x = 0,t),
\end{equation} 
where we used that there is no flux across the boundary at $x = \ell_p$. Thus, the added or extracted mass is given by 
\begin{align}
\label{mdptime}
m_{dp} = m_i + w \int\limits_0^\infty dt u_{dp}(x = 0,t) c(x = 0,t). 
\end{align}
In the following, we first determine the diffusiophoretic drift, then we deal with the cases of extraction ($\Gamma_p > 0$) and 
addition of particles  ($\Gamma_p < 0$) separately. 

% Furthermore, we assume that the particle concentration at $x = 0$ is
% constant and equal to $c_i$, that is, we do not account for depletion of
% particles from the transmitting pore, which is an issue for strong
% diffusiophoretic drift into the dead-end pore.

%%%%%%%%%%%%%%%%%%%%%%%%%%%%%%%%%%
\subsection{Diffusiophoretic drift}
%%%%%%%%%%%%%%%%%%%%%%%%%%%%%%%%%%
We focus here on estimating the drift $u_{dp}$. The P\'eclet number for salt is so low that we can
assume that diffusion dominates in the DEP. Thus, to obtain the salt
concentration $s$, we solve the diffusion equation
\begin{align}
\frac{\partial s}{\partial t} - D_s \frac{\partial^2 s}{\partial x^2} = 0. 
\end{align}
We consider the boundary conditions $s = s_H$ at $x = 0$ and $\partial s /
\partial x = 0$ at $x = L$. The initial condition is $s(x, t = 0) = s_i$. In
Laplace space we obtain the exact solution 
\begin{align}
s^\ast(x,\sigma) = \frac{s_i}{\sigma} + \frac{(s_H - s_i)}{\sigma}
\frac{\cosh[(1 - x/\ell_p) \sqrt{\sigma \tau_{D_s}}]}{\cosh(\sqrt{\sigma \tau_{D_s}})},
\end{align}
where $\tau_{D_s} = \ell_p^2/D_s$ and $\sigma$ is the Laplace variable. The
Laplace transform is defined in~\cite{AS1972}.

The Laplace transform of the diffusiophoretic velocity at $x = 0$ is then given by
\begin{align}
u_{dp}^\ast(x = 0,\sigma) = -\Gamma_p (1 - \chi) \frac{\tanh(\sqrt{\sigma
    \tau_{D_s}})}{\sqrt{\sigma D_s}},  
\end{align}
where we defined $\chi = s_i / s_H$. The integral of the drift from $t = 0$ to $\infty$ is given by 
\begin{align}
\label{intudpexact}
\int\limits_0^\infty dt u_{dp}(x = 0,t) =  u_{dp}^\ast(x = 0,\sigma = 0) = - \frac{\Gamma_p (1 - \chi) \ell_p}{D_s}. 
\end{align}
 This expression is used directly to estimate the total mass of trapped
  particles for the extraction case, as argued below. For for the trapping
  case, however, the full time dependence of $u_{dp}(,0,t)$ is required as can
  be seen from Eq.~\eqref{mdptime}.
Thus, we approximate the salt concentration
profile in the DEP by the
solution for a semi-infinite domain,
\begin{equation}
s^a(x,t)=s_i + (s_H-s_i)\text{erfc}(x/\sqrt{4D_s t}) \label{s_ana},
\end{equation}
where the superscript $a$ denotes approximation. 
Using this expression, the diffusiophoretic drift is given by
\begin{equation}
\label{udp_approx}
u^a_{dp}(x,t) =  - \Gamma_p (s_H-s_i) \frac{\exp{(-x^2/4D_s t)}}{s(x,t)\sqrt{\pi D_s t}}.
\end{equation}
 The drift at $x = 0$ then is given by 
\begin{equation}
u^a_{dp}(x = 0,t) =  - \frac{\Gamma_p (1 - \chi) }{\sqrt{\pi D_s t}},
\end{equation}
where we used that $s(x=0,t) = s_H$. For times larger than $\tau_{D_s} =
\ell_p^2/D_s$, the salt gradient decays exponentially fast with time. Thus, the
time integral over the drift can be written as
\begin{align}
\int\limits_0^\infty dt u_{dp}(x = 0,t) = \int\limits_0^{\tau_{D_s} a} dt
u^a_{dp}(x = 0,t) = - \sqrt{\frac{4}{\pi a}} \frac{\Gamma_p (1 - \chi) \ell_p}{\sqrt{D_s}}. 
\end{align}
In order to match the exact expression~\eqref{intudpexact}, we set $a = \pi/4$
and use the following the approximation 
\begin{align}
\label{udpa}
u^a_{dp}(x = 0,t) = - \frac{\Gamma_p (1 - \chi) }{\sqrt{4 \pi D_s t}}
H(\tau_{D_s} \pi - t),
\end{align}
where $H(t)$ denotes the Heaviside step function.  

%%%%%%%%%%%%%%%%%%%%%%%%%%%%%%%%%%%
\subsection{Extraction of particles}
%%%%%%%%%%%%%%%%%%%%%%%%%%%%%%%%%%%
In the case $\Gamma_p > 0$, particles are extracted from the DEP. The
particle concentration at $x = 0$, that is, at the interface 
with the TP is set equal to $c(x = 0,t) = c_i$, the resident particle concentration. 
Thus, we obtain by integration of Eq.~\eqref{eq:mdp} for the added particle mass
\begin{align}
\label{eq:mdp2}
m_{dp} = m_i + c_i w \int\limits_0^\infty dt u_{dp}(x = 0,t) =
m_i + c_i w  u_{dp}^\ast(x = 0,\sigma = 0) = m_i - \frac{m_i \Gamma_p (1 - \chi)}{D_s}, 
\end{align}
where $m_i = c_i w \ell_p$ is the initial particle mass and
$\chi=s_i/s_H$. Note that we used expression~\eqref{intudpexact} to
arrive at this result. If $\alpha_0$ is the fraction of particle mass inside the DEP
without DP, then the fraction $\alpha$ of particles after DP is 
\begin{align}
\alpha = \alpha_0 \frac{m_{dp}}{m_i}. 
\end{align}
Using Eq.~\eqref{eq:mdp2} and setting $\ell_p = \lambda$, we obtain
\begin{equation}
\alpha = \alpha_0 \left[1 - \Gamma_p^\ast Pe_s (1 - \chi) \right],
\label{alpha_mp}
\end{equation}
where $\Gamma_p^*=\Gamma_p/\lambda U$ is the dimensionless form of the
diffusiophoretic mobility and $Pe_s = \lambda U/D_s$ is the salt P\'eclet
number. 
%%%%%%%%%%%%%%%%%%%%%%%%%%%%%%%%%%%
\subsection{Trapping of particles}
%%%%%%%%%%%%%%%%%%%%%%%%%%%%%%%%%%%
In the case $\Gamma_p < 0$, particles are trapped from the TP into the
DEP. In order to determine $c(x = 0,t)$, we consider the balance of
fluxes across the interface. For $x < 0$, that is
within the TP, the particle flux transverse to the flow direction
is due to diffusion. For $x > 0$, that is, in the DEP the particle
flux is dominated by the diffusiophoretic drift. Thus, we can write
\begin{align}
\label{flux_balance}
- D_p \frac{c(x=0,t) - c_i}{\ell_0} = u_{dp}(x = 0,t) c(x = 0,t),
\end{align}
where $\ell_0$ is the concentration gradient scale. We assume that the
particle concentration in the flow past the interface
between TP and DEP is constant and equal to the initial concentration
$c_i$. We estimate $\ell_0$ as the length scale at which the diffusive flux
transverse to the flow direction toward the interface is of the same
order as the advective flux past the interface, that is,
\begin{align}
%\overline u c_i \sim D_p \frac{c_i}{\ell_0}. 
U c_i \sim D_p \frac{c_i}{\ell_0}. 
\end{align}
From this relation, we obtain the estimate
\begin{align}
%\ell_0 \sim \frac{D_p}{\overline u} = \frac{\ell_p}{Pe_p}. 
\ell_0 \sim \frac{D_p}U = \frac{\ell_p}{Pe_p}. 
\end{align}
That is, particles within the layer of thickness $\ell_0$ are available for
trapping in the DEP. \\
From~\eqref{flux_balance}, we obtain for $c_0(t) \equiv c(x = 0,t)$
\begin{align}
\label{c0}
c_0(t) = \frac{c_i}{1 + \frac{u_{0}(t) \ell_0}{D_p}},
\end{align}
where we set $u_0(t) = u_{dp}(x=0,t)$. Inserting~\eqref{c0} into~\eqref{mdptime} gives 
\begin{align}
\label{mdptrap}
m_{dp} = m_i + w \int\limits_0^\infty dt u_{0}(t) \frac{c_i}{1 + \frac{u_{0}(t) \ell_0}{D_p}}. 
\end{align}
Note that here use the approximation~\eqref{udpa} for $u_0(t)$ to
derive an analytical expression for $m_{dp}$. 
%
%%%%%%%%%%%%%%%%%%%%%%%%%%%%%%%%%%%%%%%%%%%%%
\begin{figure}
\includegraphics[width=0.7\textwidth]{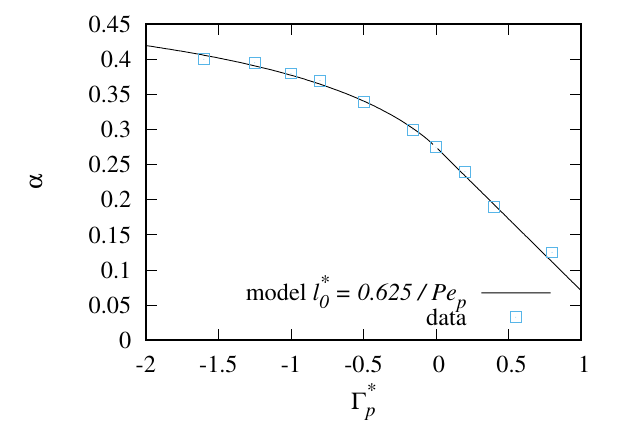}
\caption{Data for the dependence of $\alpha$ on $\Gamma_p^\ast$ (symbols) and
  the analytical model (solid line) for  $Pe_s = 0.75$, $Pe_p = 750$ and $\ell_0^\ast = 0.65/Pe_p $. }
\end{figure}
%%%%%%%%%%%%%%%%%%%%%%%%%%%%%%%%%%%%%%%%%%%%%
%

Inserting expression~\eqref{udpa}
into the right side of Eq.~\eqref{mdptrap} gives 
\begin{align}
m_{dp} = m_i + w  \int\limits_0^{\tau_{D_s} \pi/4} dt \frac{\hat \Gamma_p
  (1-\chi)}{\sqrt{\pi D_s t}} \frac{c_i}{1 + \frac{\hat \Gamma_p (1-\chi)
    \ell_0}{\sqrt{\pi D_s t}D_p}}. 
\end{align}
where we set $\hat \Gamma_p = - \Gamma_p$. We can further write
\begin{align}
m_{dp} = m_i + w c_i  \int\limits_0^{\tau_{D_s} \pi/4} dt \frac{\hat \Gamma_p
  (1-\chi)}{\sqrt{\pi D_s t} + \frac{\hat \Gamma_p (1-\chi) \ell_0}{D_p}}. 
\end{align}
Integration of the latter gives 
\begin{align}
m_{dp} = m_i + w \ell_p c_i \left\{\frac{\hat\Gamma_p (1-\chi)}{D_s} +
\frac{2 \hat \Gamma_p^2 (1 - \chi)^2 \ell_0}{\pi D_s D_p \ell_p} 
\ln\left[\frac{2 \hat \Gamma_p (1 - \chi)}{2 \hat \Gamma_p (1 - \chi) + \pi D_p \ell_p / \ell_0} \right] \right\}
\end{align}
Thus, we obtain for $\alpha$
\begin{align}
\alpha = \alpha _0 \left\{1 + \frac{\hat\Gamma_p (1-\chi)}{D_s} +
\frac{2 \hat \Gamma_p^2 (1 - \chi)^2 \ell_0}{\pi D_s D_p \ell_p} 
\ln\left[\frac{2 \hat \Gamma_p (1 - \chi)}{2 \hat \Gamma_p (1 - \chi) + \pi D_p \ell_p / \ell_0} \right]\right\}
\end{align}
We set $\ell_p = \lambda$ and define the dimensionless diffusiophoretic mobility and the dimensionless
diffusion layer scale as
\begin{align}
\label{eq:gamma}
\Gamma^\ast_p = - \frac{\hat \Gamma_p}{\lambda U}, && % \beta  = \frac{D_p}{D_s}
% \frac{\lambda}{\ell_0} = \frac{Pe_s}{Pe_{p}} \frac{1}{\ell_0^\ast} &&
\ell_0^\ast = \frac{\ell_0}{\lambda} \sim 1/Pe_p.  
\end{align}
Thus, we can write expression~\eqref{eq:gamma} in dimensionless form as
\begin{align}
\alpha = \alpha _0 \left\{1 - \Gamma_p^\ast (1-\chi)Pe_s + \frac{{2 \Gamma_p^\ast}^2
  (1 - \chi)^2 Pe_s Pe_{p} \ell_0^\ast}{\pi} 
\ln\left[\frac{2 \Gamma_p^\ast (1 - \chi) Pe_{p} \ell_0^\ast}{2 \Gamma_p^\ast (1 - \chi) Pe_{p} \ell_0^\ast- \pi
 } \right] \right\} .
\end{align}

\bibliographystyle{unsrt}
\bibliography{ref}% Produces the bibliography via BibTeX.

%======================================================